# Single indium atoms and few-atom indium clusters anchored onto graphene via silicon heteroatoms

Kenan Elibol,[1,2,3] Clemens Mangler,[1] David D. O'Regan,[2,4] Kimmo Mustonen,[1] Dominik Eder,[5] Jannik C. Meyer,[1,6] Jani Kotakoski,[1] Richard G. Hobbs,[2,3] Toma Susi,[1,*] Bernhard C. Bayer[1,5,*]

[1]Faculty of Physics, University of Vienna, Boltzmanngasse 5, A-1090, Vienna, Austria

[2]Centre for Research on Adaptive Nanostructures and Nanodevices (CRANN) and the SFI Advanced Materials and Bio-Engineering Research Centre (AMBER), Dublin 2, Ireland

[3]School of Chemistry, Trinity College Dublin, The University of Dublin, Dublin 2, Ireland

[4]School of Physics, Trinity College Dublin, The University of Dublin, Dublin 2, Ireland

[5]Institute of Materials Chemistry, Vienna University of Technology (TU Wien), Getreidemarkt 9/165, A-1060 Vienna, Austria

[6]Institute for Applied Physics, University of Tübingen, Auf der Morgenstelle 10, 72076 Tübingen

*Corresponding authors: bernhard.bayer-skoff@tuwien.ac.at, toma.susi@univie.ac.at

# Abstract

Single atoms and few-atom nanoclusters are of high interest in catalysis and plasmonics, but pathways for their fabrication and stable placement remain scarce. We report here the self-assembly of room-temperature-stable single indium (In) atoms and few-atom In clusters (2-6 atoms) that are anchored to substitutional silicon (Si) impurity atoms in suspended monolayer graphene membranes. Using atomically resolved scanning transmission electron microscopy (STEM), we find that the exact atomic arrangements of the In atoms depend strongly on the original coordination of the Si "anchors" in the graphene lattice: Single In atoms and In clusters with 3-fold symmetry readily form on 3-fold coordinated Si atoms, whereas 4-fold symmetric clusters are found attached to 4-fold coordinated Si atoms. All structures are produced by our fabrication route without the requirement for electron-beam induced materials modification. In turn, when activated by electron beam irradiation in the STEM, we observe *in situ* the formation, restructuring and translation dynamics of the Si-anchored In structures: Hexagon-centered 4-fold symmetric In clusters can (reversibly) transform into In chains or In dimers, whereas C-centered 3-fold symmetric In clusters can move along the zig-zag direction of the graphene lattice due to the migration of Si atoms during electron-beam irradiation, or transform to Si-anchored single In atoms. Our results provide a novel framework for the controlled self-assembly and heteroatomic anchoring of single atoms and few-atom clusters on graphene.

# Introduction

Supported single atoms and atomic clusters comprised of only few atoms ("nanoclusters") have distinct electrical, optical, magnetic and catalytic properties.[1–6] These have suggested single atoms and few-atom clusters to be of high application potential, particularly in heterogeneous catalysis ("single-atom catalysts" to "few-atom catalysts")[1,2,7–16] as well as in nanoplasmonics.[17,18] For catalysis applications, single atoms and few-atom clusters anchored to carbon substrates are of particularly high relevance.[16,19–24] Shortcomings in fabrication and synthesis, controlled placement and anchoring, and characterisation in terms of, e.g., structure, composition and stability, currently hinder however their further study and use.

An emerging pathway towards both controllable fabrication and concurrent microscopic imaging of small atomic arrangements is the application of aberration-corrected scanning transmission electron microscopy (STEM) to atoms that are placed onto or implanted into suspended two-dimensional (2D) membranes such as graphene.[25–29] Controlled placement of individual impurity atoms within the graphene lattice via the controlled movement of the STEM electron beam (e-beam) has been demonstrated,[30–32] mostly relying on ubiquitous substitutional Si impurities inherent to chemical vapour deposited (CVD) graphene.[33] This STEM-based manipulation has been expanded towards the study of homoatomic Si clusters in graphene.[34–38] Recently, the merging of such Si clusters with a single Pt atom towards heteroatomic Pt-Si structures on graphene was also reported.[39] A shortcoming of the STEM manipulation approach is, however, the difficulty in extending it to large scale fabrication. A materials system that intrinsically and readily allows single-atom and few-atom cluster formation and anchoring on graphene membranes in a more scalable and self-assembled fashion has remained elusive to date.

We report here the self-assembly of room-temperature-stable single In atoms (1 atom) and few-atom In clusters (2-6 atoms) and their concurrent anchoring onto individual substitutional Si dopant atoms in graphene membranes. We employ atomically resolved and element-sensitive STEM[40] to observe the In-Si structures including probing their structural dynamics and formation mechanisms *in situ* via e-beam induced structural rearrangements.[41,42] We find that the original coordination of Si in the graphene lattice is critical for the atomic arrangement of the In single atoms and clusters. While single In atoms and 3-fold symmetric

In clusters form on 3-fold coordinated Si atoms, 4-fold symmetric clusters are found on 4-fold coordinated Si atoms. Within the 4-fold and 3-fold symmetric clusters, In atoms can be either located on top of carbon atoms or hexagon centers. Hexagon-centered 4-fold symmetric $In_6$ clusters can dynamically transform into three different structures during e-beam irradiation: a chain with six In atoms, an In dimer located on a pentagon center, and a dimer located on a hexagon center. Unlike the 4-fold symmetric clusters which are immobile under electron irradiation,[33] the C-centered 3-fold symmetric $In_5$ clusters can move within the graphene lattice due to the electron irradiation induced movement of the underlying 3-fold coordinated Si site. These 3-fold clusters can also transform to anchored single In atoms.

Combined, our observations evidence a large diversity of room-temperature-stable single atoms and few atom clusters in the In-Si-graphene system and thereby provide a unique materials system and unprecedented toolbox for the controlled self-assembly and heteroatomic anchoring of single atoms and few-atom clusters onto graphene.

# Results

The experiments we carried out to deposit indium single atoms and few-atom clusters on graphene are schematically shown in Figure 1a. A monolayer CVD graphene membrane suspended over a holey SiN chip is first loaded into the STEM setup (ultra-high vacuum (UHV) with base pressure ~$10^{-9}$ mbar) which comprises the microscope and directly coupled preparation chambers including laser annealing and an *in situ* evaporation chamber. We note that CVD graphene intrinsically includes a small number of substitutional Si heteroatoms, as well-documented in prior literature.[33,43,44] The sample is then first irradiated in UHV by 600 mW of laser power at 445 nm wavelength, allowing the removal of adventitious hydrocarbon adsorbates from the graphene that stem from membrane fabrication and ambient air exposure.[45,46] Subsequently, without vacuum break, In is evaporated onto the graphene membrane using a custom-built preparation chamber (base pressure ~$10^{-9}$ mbar).[42] After *in situ* deposition of In, the sample is transferred into the microscope, without vacuum break, and once more irradiated *in situ* by a similar 600 mW laser before STEM imaging. Due to the low melting point (~160 °C) but low vapor pressure of In,[47] the second laser irradiation critically promotes the diffusion of In atoms over the graphene surface.

During STEM imaging, medium and high angle annular dark field (MAADF and HAADF) signals are acquired simultaneously. These allow the discrimination of imaged elements based on signal intensity (our HAADF intensity scales with atomic number as $Z^{~1.6}$ for isolated atoms and roughly linear with specimen thickness for a given $Z$).[40] Further elemental identification is made via simultaneously recorded electron energy loss spectroscopy (EELS) data. Further experimental details can be found in the Methods section.

Figure 1b-f shows that large atomically clean areas are found on the graphene after In deposition and the second laser irradiation. The key feature of interest we observe in these large clean graphene areas are the small bright spots in the MAADF images in Figure 1b-f. We identify these spots as single In atoms (Figure 1c) and few-atom In clusters formed on the graphene surface (Figure 1e,f), which are all anchored to the graphene via single substitutional Si impurity atoms covalently bound within the graphene lattice, as described in detail below. Notably for the clusters, we consistently observe two different symmetries: 3-fold and 4-fold. (These symmetries refer to as-observed contrast within the image projection plane, as typical for (S)TEM measurements. Out-of-plane atom positions in the structures are not readily accessible by STEM, but will be elucidated based on our associated structural modelling and image simulations).

We stress that the single In atoms and In clusters are already present before extended STEM imaging, i.e. they are generally not produced by the electron irradiation, but have self-assembled during In evaporation and the second laser anneal prior to imaging. In the fields of view in Figure 1b,d (with electron irradiation dose rates of $0.73 \times 10^6$ e$^-$nm$^{-2}$s$^{-1}$ (b), $0.18 \times 10^6$ e$^-$nm$^{-2}$s$^{-1}$ (d)), the In structures appear to be unperturbed by STEM imaging. We verify the structure of these In clusters on graphene at higher magnification in the next sections. Imaging at higher magnifications is associated with higher electron dose rates, however, which drives structural modifications of the In structures by energy transfer from the scanning e-beam,[41,42] but at the same time also allows us to *in situ* probe their structural dynamics.

To verify the chemical nature of the structures, an EEL spectrum image is acquired on a typical cluster shown in the HAADF image of Figure 2a. EELS accumulated with an energy range focused on the In *M*- and Si *L*-edges show that the cluster consists of In and Si atoms (Figure 2b,c). The EELS data indicates 4-fold symmetry for the positions of In around a rather central Si signal. This is a first hint to the Si-anchoring of the In atoms on the graphene lattice. From the EELS map, however, it remains unclear if the Si signal is from one or several central Si atoms.

Towards addressing this question, the HAADF image sequence in Figure 2d-f (see also Supplementary Video 1) reveals the self-assembly process of the Si-In structures (here emulated by the energy input from the e-beam). This sequence has been acquired at a higher magnification (and thus higher electron dose rate of $20.83 \times 10^6$ e$^-$nm$^{-2}$s$^{-1}$) than Figure 1b,d and thereby enables observation of e-beam induced structural dynamics.[41,42] In particular, Figure 2d shows initially a bare 4-fold coordinated Si atom in the graphene lattice. After 8.3 s (electron dose $0.1 \times 10^9$ e$^-$nm$^{-2}$), In atoms have become trapped at this Si atom to form an intermediate Si-anchored In structure (Figure 2e, further details on this intermediate possible dimer structure are discussed below in Figure 4). After 25.1 s (Figure 2f, electron dose $0.29 \times 10^9$ e/nm$^2$), additional In atoms have become trapped to eventually form a 4-fold symmetric In cluster anchored to the 4-fold Si atom in the graphene lattice (further details on the structure below).

This observation of In cluster formation suggests that a single central Si atom is present in the observed In cluster structures. Notably, no In reservoir is present within the field of view around the Si atom during cluster formation in Figure 2d-f. This excludes a direct role of the electron beam in moving the In atoms from an reservoir towards the Si during STEM scanning.[48] Instead, Figure 2d-f suggests a self-assembly formation mechanism for the In clusters in which In atoms or moieties constantly diffuse across the graphene membrane (with much higher diffusion speed than resolvable by STEM[41,49]), and become trapped during encounters with the substitutional Si, thereby self-assembling into the observed Si-anchored In clusters on the graphene. We suggest that this facile self-assembly relies on the significant mobility[50] of In atoms on the graphene membrane at room temperature (resulting from In's low melting point (~160 °C) and low vapor pressure over a wide temperature range[47]) and the higher reactivity of the Si site compared to the relatively inert graphene lattice.

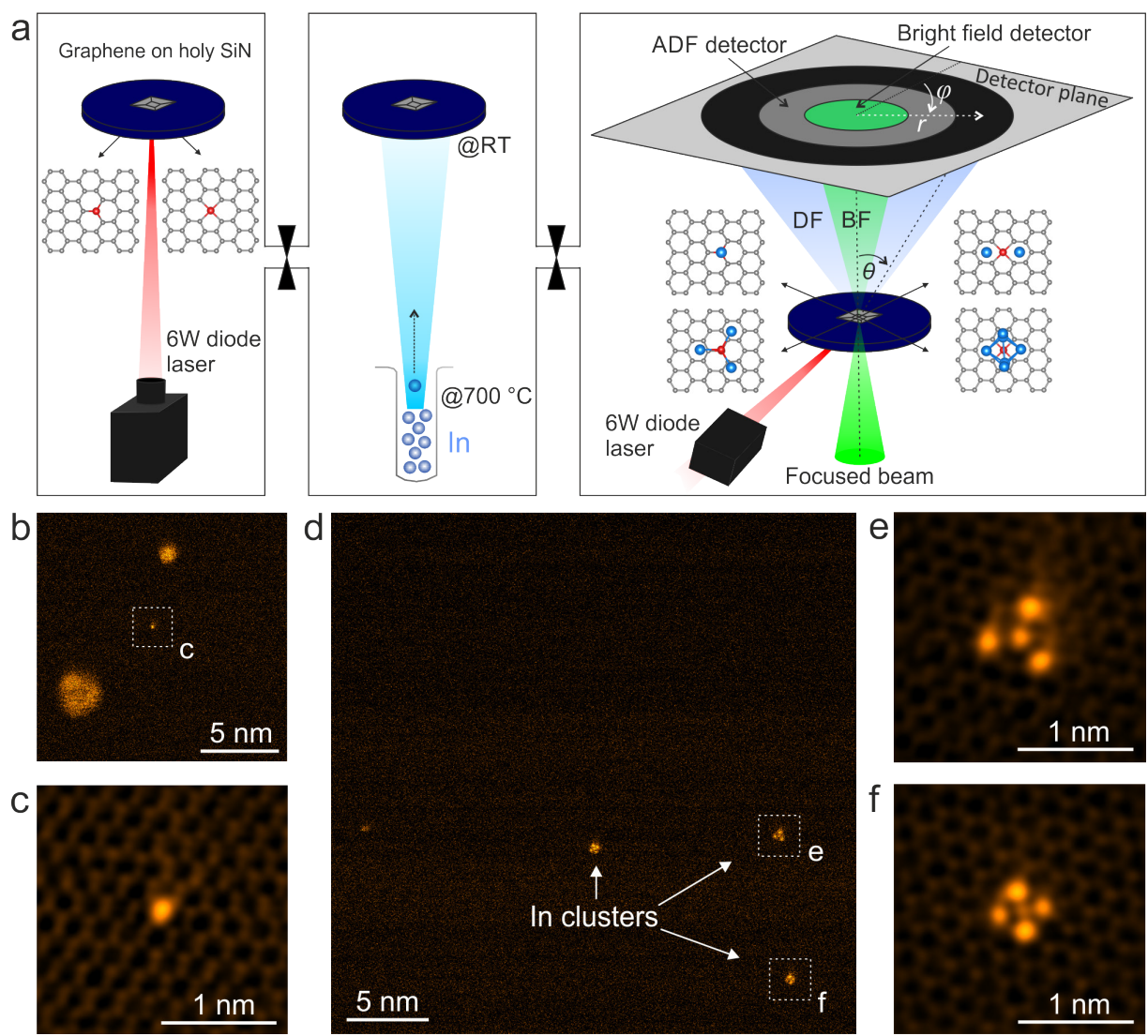

**Figure 1. Overview of In single atom and cluster anchoring on graphene.** (a) Schematic illustration of the experiment involving laser irradiation, *in situ* In deposition, and imaging by STEM. The blue, red and grey coloured atoms are In, Si and C, respectively. (b,d) Large-area MAADF-STEM images of a clean monolayer graphene surface after *in situ* deposition, showing single In atom (b) and few-atom In clusters (d). (c,e,f) Close-up MAADF-STEM images of a single In atom and 3-fold and 4-fold symmetric In clusters indicated by white dashed frames in (b,d). All images display false colour and images in (c,e,f) are double Gaussian filtered after Wiener filtering to reduce noise and enhance contrast (see Methods section; raw images are shown in Supplementary Figure 1). Note that the contrast of the underlying graphene lattice is low compared to the clusters, making it difficult to display both clearly in some of our figures.

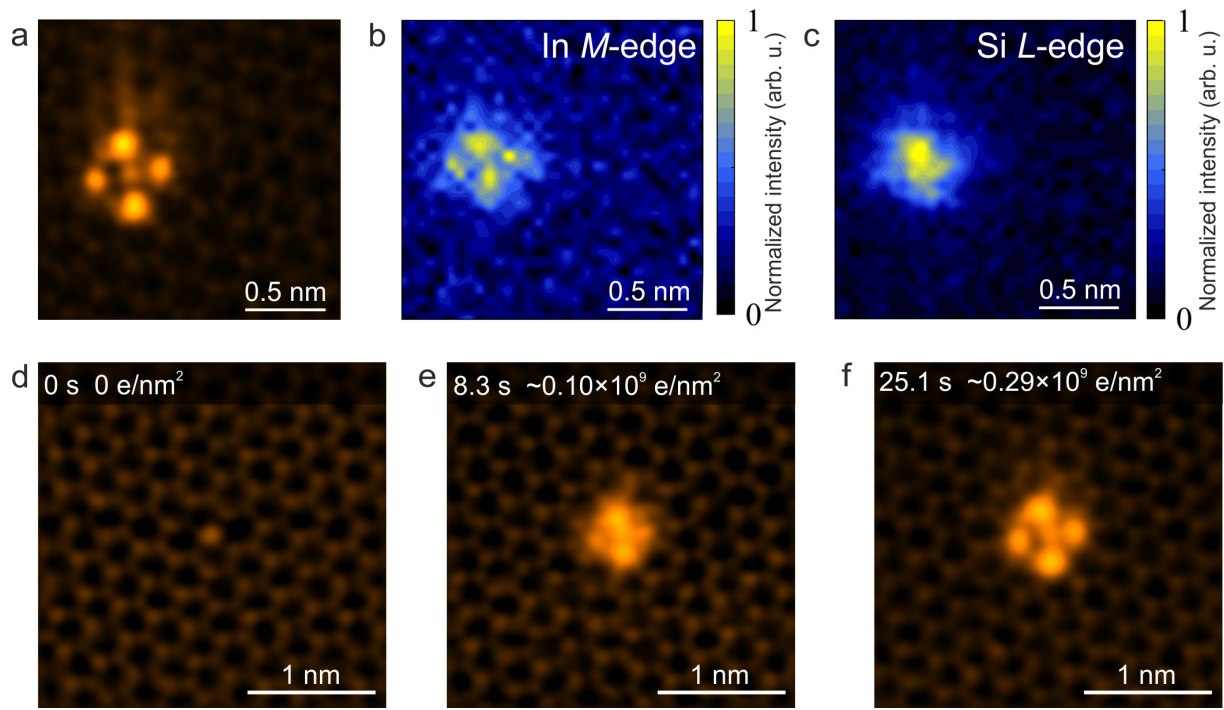

**Figure 2. Trapping of In on substitutional Si impurities in the graphene lattice.** (a) HAADF-STEM image of a 4-fold symmetric cluster and (b-c) its simultaneously acquired EEL spectrum maps showing the In *M*-edge (b) and the Si *L*-edge (c), respectively. (d-f) MAADF-STEM images of (d) 4-fold coordinated Si impurity in graphene, (e) intermediate In structure anchored onto this Si atom and (f) 4-fold symmetric In cluster trapped on the same Si atom. All experimental HAADF and MAADF images are in false colour and filtered (see Methods section; raw images are shown in Supplementary Figure 2). For EEL spectra of In atoms see Supplementary Figure 3.

After this first assessment of the overall nature of the anchored clusters, we now turn to detailed atomic structure analysis aided by density functional theory (DFT)-based structure modelling. In Figure 3, the 4-fold symmetric In clusters from Figure 1f and Figure 2 are analysed: Our data reveals that there are two different 4-fold symmetric In clusters forming on 4-fold coordinated Si atoms as shown in Figure 3a,b. Based on the EELS findings and the cluster formation sequence in Figure 2, we have relaxed atomistic models comprising of In, Si and C atoms with DFT to use in STEM image simulations (see Methods section). The first model in Figure 3c addresses the experimental observation that two of the In atomic columns are brighter than the other two, and thus involves six In atoms anchored on a 4-fold coordinated Si impurity in graphene. In this model, the In atoms are located over the hexagon centers of graphene. We thus term this type of structure as hexagon-centered 4-fold $In_6$ cluster. The image simulation shown in Figure 3e is in a good agreement with the experimental image (Figure 3a), as shown by the line profiles recorded over the experimental and simulated HAADF images (Figure 3g). (The intensity at the center of cluster is slightly higher in the experimental data, which we attribute to the probe tails that can cause intensity enhancement at the center surrounded by six intensely scattering In atoms.[51] However, note that the projected atomic positions are quite sensitive to the atoms present in the cluster.)

The second observed structure is also 4-fold symmetric (Figure 3b), but the experimental contrast is equal for all four In columns, indicating that there are four In atoms in the cluster all located over C atoms (see Figure 3d). We thus call this structure a C-centered 4-fold $In_4$ cluster. The DFT-relaxed model matches the experimental data well (see Figure 3b,f,h) (except the intensity at the center of cluster in experimental data, which is again higher compared to simulated data). Besides In atoms, we have tried atomic models including Ca, Cu and Fe metal atoms and compared the intensities and symmetries of experimental and simulated HAADF images based on DFT-relaxed models (see Supplementary Figure 4). Those result in a much poorer match, further corroborating that the bright spots on HAADF images are In. Besides these 4-fold symmetric In clusters, we considered another In structure with four In and two Si atoms (see Supplementary Figure 5). This structure is similar to that shown in Figure 3b, but the cluster contains one additional Si, which pushes the four In atoms away from the center and thus despite increasing the intensity of the central spot does not match the overall experimental contrast.

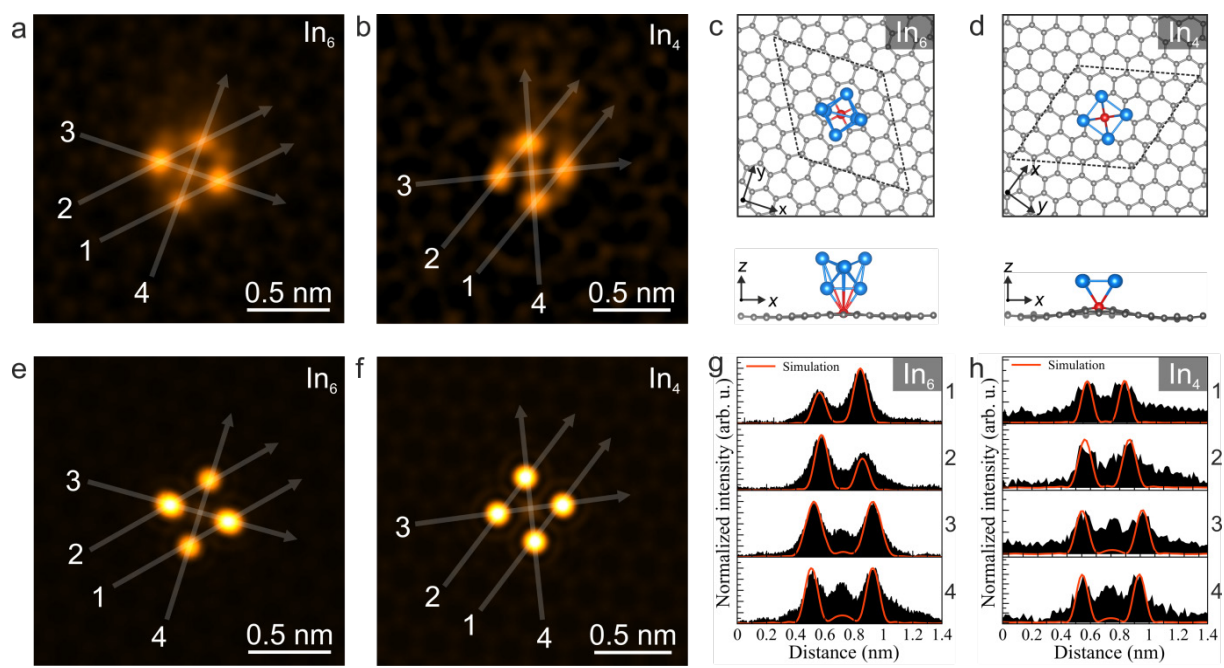

**Figure 3. Atomic structure analysis of 4-fold symmetric In clusters.** (a,b) HAADF-STEM images of hexagon-centered 4-fold symmetric $In_6$ and C-centered 4-fold symmetric $In_4$ clusters, respectively. (c,d) Plan and side views of DFT-relaxed models with six and four In atoms used in the image simulations. The dashed lines on the atomic models delineate the computational graphene 5×5 supercell containing the In cluster. The blue, red and gray coloured atoms are In, Si and C, respectively. (e,f) Simulated HAADF images corresponding to the clusters shown in (a) and (b), respectively. (g,h) Intensity profiles measured along the white dashed lines on the experimental and simulated HAADF images of 4-fold symmetric $In_6$ and $In_4$ clusters, respectively. All experimental HAADF images are shown in false colour and are double Gaussian filtered (see Methods section).

The dynamics of a hexagon-centered 4-fold $In_6$ cluster under electron irradiation (dose rate $46.88\times10^6$ $e^-nm^{-2}s^{-1}$) are shown in Figure 4a (raw images in Supplementary Video 2). In Figure 4b,c, the simulated HAADF images and atomic models matching with the experimental HAADF images of the cluster are presented. The initial structure of this cluster with six In atoms is transformed after 16.8 s (electron dose of $\sim0.79\times10^9$ $e^-nm^{-2}$) into an In chain where the In atoms are aligned along the armchair direction of the graphene lattice, with all six In atoms retained. After a further 42.0 s ($\sim1.96\times10^9$ $e^-nm^{-2}$), all In atoms of the cluster disappear from the field of view. Notably, a central 4-fold symmetric substitutional Si impurity in the graphene lattice remains visible, again corroborating the role of substitutional Si as a central "anchor" for the In clusters on graphene. After 100.7 s ($\sim4.72\times10^9$ $e^-nm^{-2}$), two

In atoms are again captured by the Si atom (similar as in the cluster formation sequence in Figure 2d-f) and located over the pentagon centers of the 4-fold Si impurity site. After 117.4 s (~$5.50\times10^9$ e⁻nm⁻²), these two In atoms rotate by 90° to overlay the graphene hexagon centers. However, they return back to their initial position by again rotating 90° after 125.8 s (~$5.90\times10^9$ e⁻nm⁻²). This structure finally attracts four additional In atoms after 167.8 s (~$7.86\times10^9$ e⁻nm⁻²) to form the hexagon-centered 4-fold $In_6$ cluster, identical to that observed in the first frame of the sequence.

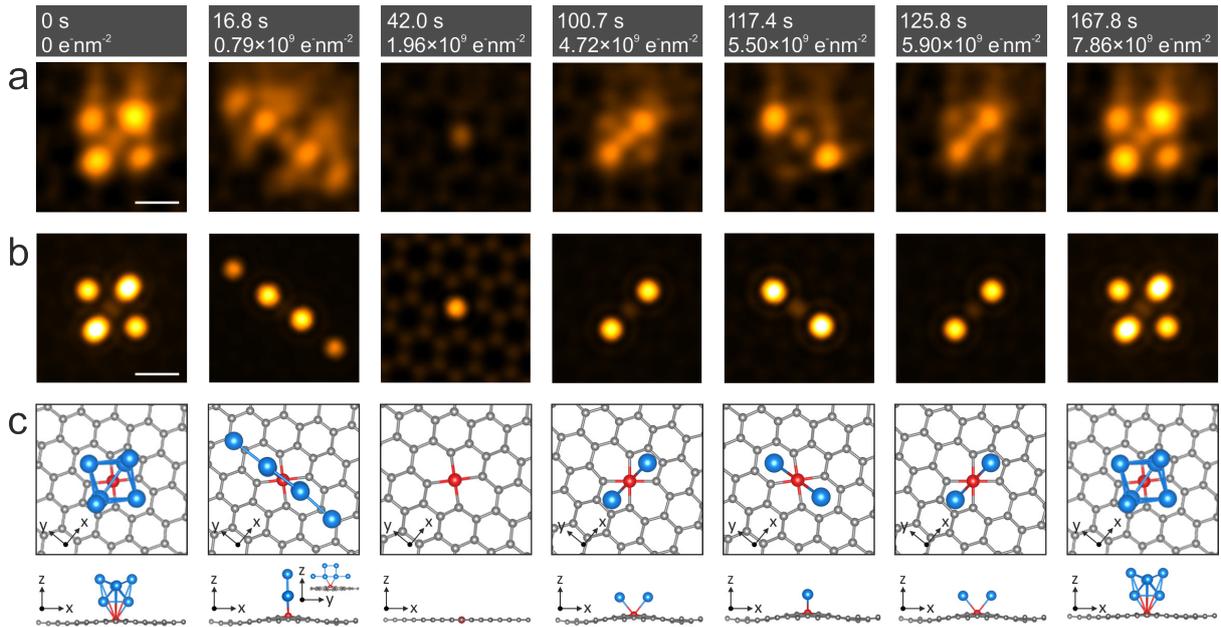

**Figure 4. Dynamics of hexagon-centered 4-fold symmetric $In_6$ cluster under e-beam exposure.** (a) HAADF-STEM images acquired at successive times, corresponding to increasing irradiation doses (indicated over the panels). (b) HAADF image simulations corresponding to the experimental images in (a). (c) Plan and side views of DFT-relaxed models with In clusters anchored on the 4-fold coordinated Si impurity in graphene. All experimental HAADF images are in false colour and double Gaussian filtered (see Methods section). The scale bars are 0.3 nm.

In addition to indium clusters formed on 4-fold coordinated Si impurities, we observe 3-fold symmetric In clusters as well as single In atoms anchored on 3-fold coordinated Si impurities, as shown in detail in Figure 5. Similar to the 4-fold symmetric cases, In atoms in 3-fold symmetric clusters may be located over the graphene C atoms or hexagon centers, and there are two distinct variants, here differing by the intensity of the central atomic column. We term these the C-centered 3-fold $In_5$ cluster (see Figure 5a,b) and the hexagon-centered 3-fold $In_3$ cluster (see Figure 5g,h). Notably the hexagon-centered 3-fold $In_3$ structure includes two Si atoms, one of which is substitutional in the graphene lattice and one which is part of the In cluster. As shown in Figure 5a-c and g-i, the DFT-relaxed models and experimental data are in a good agreement.

Under e-beam exposure at higher magnifications, the In atoms in these 3-fold clusters can be readily ejected by the focused electron beam (see Figure 5a,d, and g,j, also Supplementary Videos 3, 4, 5 and 6). HAADF-STEM images in Figure 5d show a single In atom anchored on the 3-fold coordinated Si impurity after the other In atoms are removed from the 3-fold $In_5$ cluster. Notably, although four In atoms are removed, the remaining single In atom keeps its position on the Si atom under electron irradiation (see Supplementary Videos 4, 5 and 6). In addition to the model shown in Figure 5b, we have created two different models with four and five In atoms and compared them with the experimental data (see Supplementary Figure 6). As shown in Supplementary Figure 6d, the atomic structure shown in Figure 5b is in a better agreement with the experimental image intensity.

Unlike C-centered 3-fold $In_5$, in the hexagon-centered 3-fold $In_3$ cluster only the anchoring Si impurity remains in the graphene lattice after removal of three In atoms (see also EEL spectra in Supplementary Figure 7). (We attempted to relax a C-centered 3-fold symmetric $In_3$ cluster, i.e. without an In or Si atom directly on top of the Si impurity, see Supplementary Figure 8, but the spacing of the cluster and the intensity of the center column do not match the experimental data.)

We note that, intermittently, also short-term capturing of single In atoms by 4-fold Si impurities is observed (Supplementary Figure 9), albeit these arrangements are never as stable as in the case of 3-fold Si anchoring. We also note that when the coordination of the Si atom in the graphene lattice is changed due to e-beam exposure, trapping of different structures on the same Si atom is observed. For instance, in Supplementary Figure 10 a 3-fold symmetric In cluster is first trapped by a 3-fold coordinated Si atom, then again removed, after which the Si atom switches into 4-fold coordination, which eventually enables the trapping of an $In_2$ dimer.

After having identified the various observed structures of the In-Si-graphene system, we can semi-quantitatively elucidate their relative formation probabilities by measuring their areal observation densities when imaged at wide fields of view (i.e. without inducing structural dynamics with a higher e-beam dose rate; the total number of observations was 43). Note that observed areal densities of bare 4-fold coordinated Si (0.82/1000 nm$^{-2}$) and bare 3-fold coordinated Si (0.89/1000 nm$^{-2}$) anchoring heteroatoms in the graphene lattice are roughly similar for our samples. In descending order we find the most often observed (thus most readily formed and, arguably, most stable) Si-anchored In structures to be as follows: hexagon-centered 4-fold symmetric In$_6$ clusters (0.37/1000 nm$^{-2}$), single In atom anchored on Si (0.32/1000 nm$^{-2}$), hexagon-centered In$_3$ 3-fold symmetric clusters (0.22/1000 nm$^{-2}$), C-centered 3-fold symmetric In$_5$ clusters (0.15/1000 nm$^{-2}$), In$_2$ dimers (0.12/1000 nm$^{-2}$), and In$_6$ chains (0.07/1000 nm$^{-2}$). These statistics underline that both In single atoms and few-atom clusters are well stabilized by the Si-graphene system.

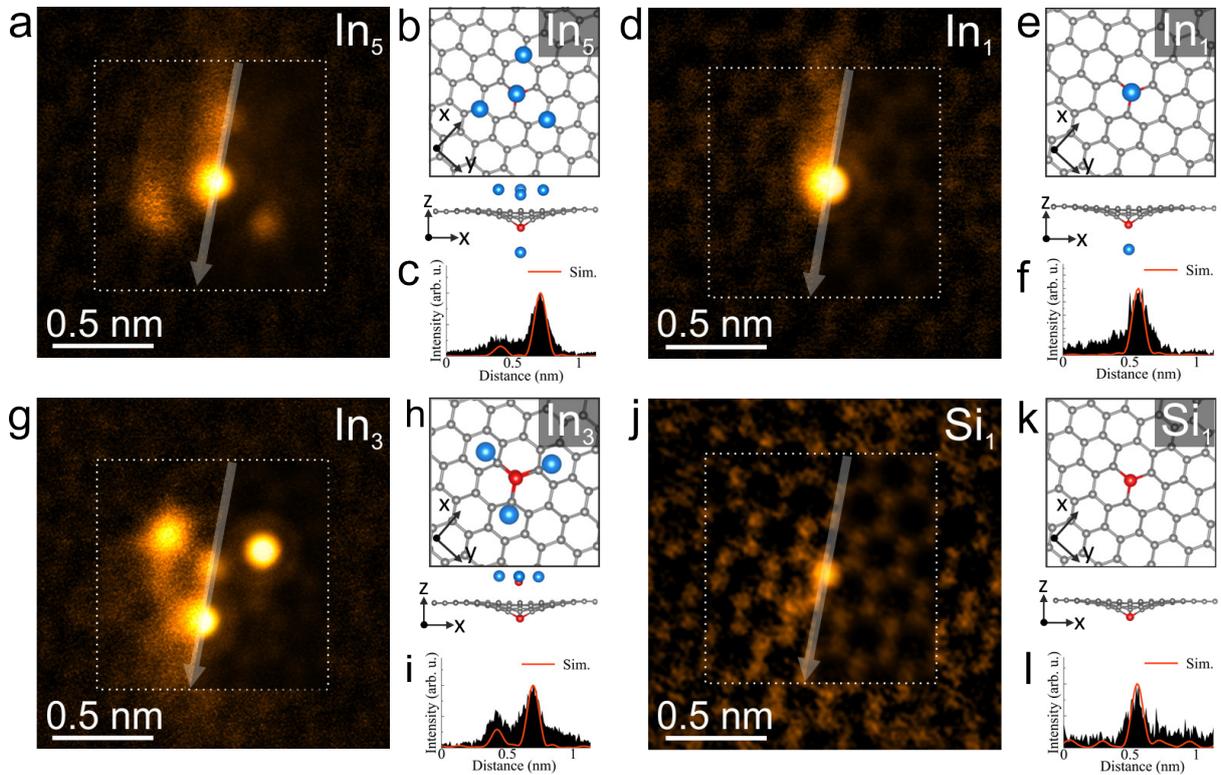

**Figure 5. Atomic structure analysis of 3-fold symmetric In clusters.** (a,d,g,j) HAADF-STEM images of (a) a C-centered 3-fold symmetric In$_5$ cluster and (d) In$_1$ single atom (the same area as in (a) after four In atoms are removed during electron irradiation at a dose of ~0.29×10$^9$ e$^-$nm$^{-2}$) as well as (g) hexagon-centered 3-fold symmetric In$_3$ cluster and (j) Si$_1$ in graphene lattice (the same area as in (g) after three In atoms surrounding the central Si atom are removed during electron irradiation at a dose of ~0.39×10$^9$ e$^-$nm$^{-2}$), respectively. Semi-transparent simulated HAADF images corresponding the structures shown in panels b, e, h

and k are shown on the right side of line profiles in white dashed frames on panels a, d, g and j (left side of line profiles display raw image). The experimental images are in false colour and Wiener filtered (raw images are shown in Supplementary Figure 11). (b,e,h,k) Plan and side views of the DFT-relaxed models used for the image simulations shown in panels a, d, g and j, respectively. The blue, red and grey coloured atoms shown in the atomic models are In, Si and C, respectively. (c,f,i,l) Intensity profiles recorded along the semi-transparent white lines overlaid on the experimental and simulated HAADF images in (a,d,g,j). The identity of the Si site was verified by EELS (Supplementary Figure 7).

When irradiated by the e-beam at higher dose rates (smaller fields of view), we not only observe cluster formation and structural dynamics *within* the In clusters (Figures 2, 4, 5) but for some also translation/migration in their entirety along the graphene lattice: While the hexagon-centered $In_3$ clusters are not observed to change their position during e-beam irradiation (see Supplementary Video 3), we show in Figure 6a the dynamics of a C-centered $In_5$ 3-fold symmetric In cluster during irradiation (see also Supplementary Video 7). From its initial position, the cluster moves after 4.2 s (~0.05×10$^9$ e$^-$nm$^{-2}$) with respect to the graphene lattice. The movement of the cluster is due to the migration of the Si impurity to a neighboring C site along the zig-zag direction of graphene, presumably via the same bond inversion mechanism as Si impurity manipulation.[33] The initial positions of the In atoms are marked by red circles in the DFT-relaxed model corresponding to the structure acquired at 4.2 s (~0.05×10$^9$ e$^-$nm$^{-2}$). (Note that the Si atom is not visible in the model due to the In atom located on top of it.) The change in the location of the Si changes the alignment of the In atoms with respect to the graphene lattice, and thus the entire cluster migrates to preserve its bonding. The movement of the cluster continues after 8.4 s (~0.10×10$^9$ e$^-$nm$^{-2}$). In the HAADF image, an additional In contrast appears below the cluster (same effect is visible in the frame at 21.0 s at a dose of ~0.25×10$^9$ e$^-$nm$^{-2}$). We believe this is because the cluster rotated before the scan was finished, which suggested acquiring STEM images at a higher scan speed. Although we were not able to resolve the rotation direction of the cluster, additional In contrast does appear on the HAADF image at 8.4 s (~0.10×10$^9$ e$^-$nm$^{-2}$) suggesting that the cluster rotates 60° anticlockwise around the center (see also Supplementary Figure 12). Until the last image frame acquired after 54.5 s (~0.64×10$^9$ e$^-$nm$^{-2}$), the Si atom and the cluster has jumped five times along the zig-zag direction of the graphene lattice (see path in Figure 6d). Although beyond the scope of this work, this observation sequence suggests that not only single atoms,[31] but also more complex heteroatomic structural assemblies could be positioned on the graphene lattice by electron-beam manipulation.

Finally, large-supercell semi-local DFT calculations[52] and their resulting atom-decomposed densities of states (Supplementary Figure 13-14) indicate that the here observed Si-In structures on graphene are all approximately one- or two-electron deficient with a large local charge polarization and, as such, may be promising candidates for application screening e.g. for single-site catalysis.

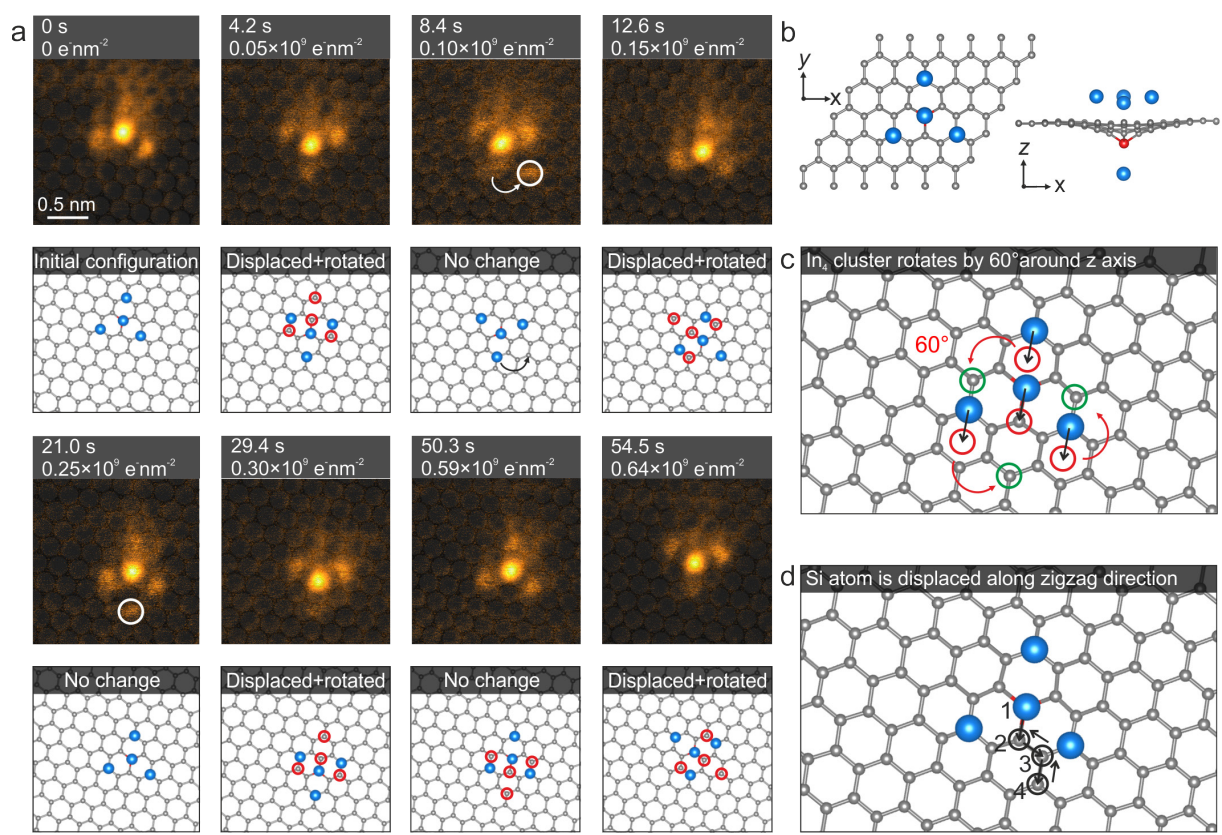

**Figure 6. Dynamics of C-centered 3-fold symmetric In$_5$ cluster under electron irradiation.** (a) MAADF-STEM image sequence of the cluster acquired at different electron irradiation doses, and the corresponding DFT-relaxed models. A semitransparent DFT model is superimposed on all images. The red circles represent the previous locations of In atoms before the cluster displaces and rotates. (b) DFT-relaxed model of C-centered In$_5$. (c) A DFT-relaxed model showing how the cluster rotates after displacement of the Si atom to a neighboring C site. The red circles indicate the position of In atoms after migration of cluster by half a lattice vector (see black arrows). The green circles show the position of three In atoms surrounding the center after the rotation of the cluster by 60°. (d) The same model as in (c), showing the initial configuration of the $In_5$ cluster and the sites that the cluster occupies under electron irradiation. Firstly, the center of the cluster moves step by step from position 1 to 4 (see black circles), and then it migrates back from position 4 to 2. The black arrows show the direction of movement. The experimental images are in false colour and Wiener filtered (raw images are shown in Supplementary Figure 12).

## Conclusion

We report the self-assembly and anchoring of single In atoms and few-atom In clusters onto substitutional Si impurity atoms in suspended monolayer graphene membranes. A variety of structure types that are stable at room temperature are found from our facile fabrication route without the requirement for e-beam induced materials modification. Most frequently observed structures are hexagon-centered 4-fold symmetric $In_6$ clusters, single In atoms anchored on Si, and hexagon-centered 3-fold symmetric $In_3$ clusters. Notably, the original coordination of the Si determines the atomic arrangements of the In structures: While single In atoms and 3-fold symmetric In clusters form on 3-fold coordinated Si impurities, 4-fold symmetric clusters are found on 4-fold coordinated Si impurities. Due to energy transfer from the scanning e-beam, in higher dose rate close-up imaging we observe *in situ* the formation, structural changes and translation dynamics of the Si-anchored In structures on graphene: The hexagon-centered $In_6$ 4-fold symmetric clusters transform into three different structures during e-beam irradiation, including In chains and dimers. Unlike the 4-fold symmetric clusters, the C-centered $In_5$ 3-fold symmetric clusters can move under e-beam irradiation along the zigzag direction of graphene lattice, and also transform to single In atoms anchored on the Si. The observed Si-anchored In structures on graphene are promising for future application screening in, e.g., catalysis. Combined, our results provide a first materials system and framework towards the controlled self-assembly and heteroatomic anchoring of single metal atoms and few-atom clusters on graphene.

# Methods

**STEM and EELS measurements.** STEM images were acquired with a Nion UltraSTEM100 operated at a 60 kV accelerating voltage in UHV (~$10^{-9}$ mbar) using concurrent high angle annular dark field (HAADF) and medium angle annular dark field (MAADF) detectors with collection angles of 80–300 mrad and 60–80 mrad, respectively. The EELS experiments were carried out by a Gatan PEELS 666 spectrometer retrofitted with an Andor iXon 897 electron-multiplying charge-coupled device camera. The energy dispersion, the beam current and the EELS collection semi-angle were 0.5-1 eV per channel, 30 pA and 35 mrad, respectively.[53] The STEM is equipped with a custom-made sample loading and transfer system to enable direct transfer of samples from various preparation chambers into the STEM without exposure to ambient.[54,55]

*In situ* **laser cleaning.** A tunable 6 W diode laser (445 nm, Lasertack GmbH) was used to clean the graphene surfaces. Laser irradiation of the sample held in a transfer arm was performed through a viewport in both STEM and UHV sample preparation chambers. In the experiments, the laser was operated at 10% duty cycle reducing the laser power to 600 mW, which does not induce structural damage in graphene but is sufficient for cleaning.[45,46]

*In situ* **In deposition.** The *in situ* evaporation of In was achieved using a custom-built preparation chamber (base pressure ~$10^{-9}$ mbar) coupled to the STEM. The evaporation source was a Knudsen cell with In pellets (99.99% purity, Kurt J. Lesker), which were heated to 700 °C. The resulting In flux was then directed at the graphene sample, which was not intentionally heated. Nominally deposited In thickness was monitored using a quartz micro balance and kept to ~10 nm. Substrates used for In deposition were commercial CVD graphene (Graphenea Inc.) suspended on perforated silicon nitride grids (Ted Pella Inc.). Subsequent to *in situ* In evaporation, a second *in situ* laser cleaning step was applied to the sample.

**STEM image simulations.** HAADF and MAADF image simulations were carried out using the QSTEM software with parameters corresponding to the experiments:[56] chromatic aberration coefficient of 1 mm, a spherical aberration coefficient of 1 μm, energy spread of 0.48 eV. HAADF and MAADF detector angle ranges are set to the experimental range of 80–300 mrad and 60–80 mrad, respectively.

**Image processing.** ADF images were processed to reduce noise and increase contrast via double Gaussian filtering[40], in some cases after applying Wiener filtering. The parameters used for double Gaussian filtering are $\sigma 1 = 0.25$, $\sigma 2 = 0.20$, weight = 0.3 (for Figure 1c,e,f), $\sigma 1 = 0.25$, $\sigma 2 = 0.22$, weight = 0.25 (for Figure 2a), $\sigma 1 = 0.25$, $\sigma 2 = 0.22$, weight = 0.25 (for Figure 2e-g), $\sigma 1 = 0.36$, $\sigma 2 = 0.26$, weight = 0.20 (for Figure 3a), $\sigma 1 = 0.32$, $\sigma 2 = 0.25$, weight = 0.15 (for Figure 3b) and $\sigma 1 = 0.35$, $\sigma 2 = 0.26$, weight = 0.30 (for Figure 4). For the images in Figure 1, Figure 5 and Figure 6, the low-pass Wiener filter[57] was applied. In addition, we used false coloring with the ImageJ lookup table "Orange Hot".

**DFT simulations.** Density functional theory (DFT) simulations were carried out using the grid-based projector-augmented wave (GPAW) software package[58] to study the properties of the supercells of monolayer graphene with Si impurities and adsorbed In clusters. The atomic structures were relaxed with the PBE functional and periodic boundary conditions (with >10 Å of vacuum in the perpendicular direction between the images) in the LCAO mode[58] with the grid spacing of 0.2 Å and a $5 \times 5 \times 1$ **k**-point mesh so that maximum forces were <0.02 eV Å$^{-1}$.[59] For details of large-supercell density of states calculations[52], see Supplementary Figures 13-14.

# Acknowledgements

K.E., C.M. and B.C.B. acknowledge support from the Austrian Research Promotion Agency (FFG) under project 860382-VISION. T.S. acknowledges funding by the European Research Council (ERC) under the European Union's Horizon 2020 research and innovation programme (grant agreement no. 756277-ATMEN). D.D.OR. acknowledges the support of Science Foundation Ireland (SFI) through The Advanced Materials and Bioengineering Research Centre (AMBER, grant 12/RC/2278 P2), and of the European Regional Development Fund (ERDF), and further acknowledges Trinity Centre for High Performance Computing and Science Foundation Ireland, for the maintenance and funding, respectively, of the Boyle (Cuimhne upgrade) cluster on which DFT calculations were performed. R.G.H acknowledges the support of Science Foundation Ireland (SFI) through The Advanced Materials and Bioengineering Research Centre (AMBER, grant 12/RC/2278_2) and the Royal Society-Science Foundation Ireland University Research Fellowship (15/RS-URF/3306).

# Author contributions

K.E. and B.C.B conceived the idea. K.E. performed the experiments and analysed the data. C.M., J.K. and J.C.M. built the setup to carry out the experiments. K.M. helped in sample preparation. D.D.OR. performed large-scale DFT simulations. D.E. and R.H. provided input to data interpretation. J.K. provided input to and T.S. led the DFT-based structure analysis. B.C.B. supervised the project. K.E., T.S. and B.C.B. drafted the manuscript with input from all authors.

SUPPLEMENTARY INFORMATION

# Single indium atoms and few-atom indium clusters anchored onto graphene via silicon heteroatoms

Kenan Elibol,[1,2,3] Clemens Mangler,[1] David D. O'Regan,[2,4] Kimmo Mustonen,[1] Dominik Eder,[5] Jannik C. Meyer,[1,6] Jani Kotakoski,[1] Richard G. Hobbs,[2,3] Toma Susi,[1,*] Bernhard C. Bayer[1,5,*]

[1]Faculty of Physics, University of Vienna, Boltzmanngasse 5, A-1090, Vienna, Austria

[2]Centre for Research on Adaptive Nanostructures and Nanodevices (CRANN) and Advanced Materials and Bio-Engineering Research Centre (AMBER), Dublin 2, Ireland

[3]School of Chemistry, Trinity College Dublin, The University of Dublin, Dublin 2, Ireland

[4]School of Physics, Trinity College Dublin, The University of Dublin, Dublin 2, Ireland

[5]Institute of Materials Chemistry, Vienna University of Technology (TU Wien), Getreidemarkt 9/165, A-1060 Vienna, Austria

[6]Institute for Applied Physics, University of Tübingen, Auf der Morgenstelle 10, 72076 Tübingen

*Corresponding authors: bernhard.bayer-skoff@tuwien.ac.at, toma.susi@univie.ac.at

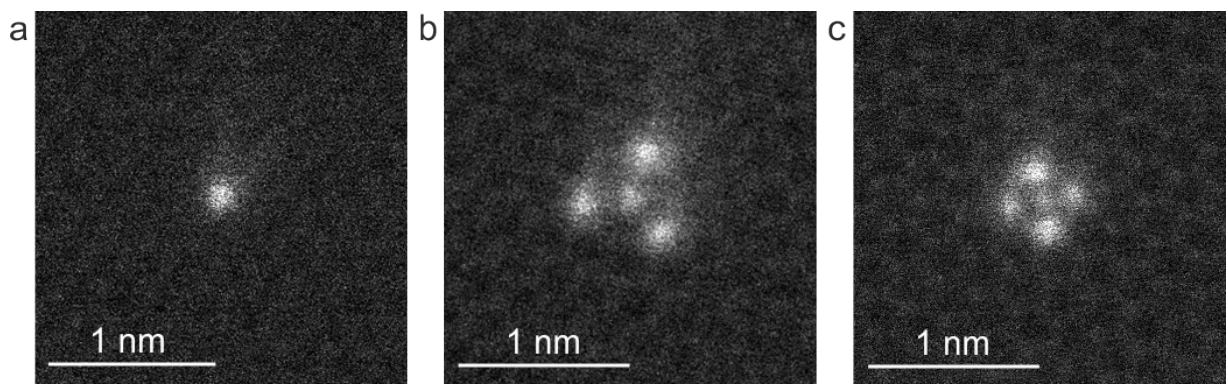

**Supplementary Figure 1.** Raw MAADF-STEM images of a single In, 3-fold and 4-fold symmetric In clusters shown in Figure 1 in the main text.

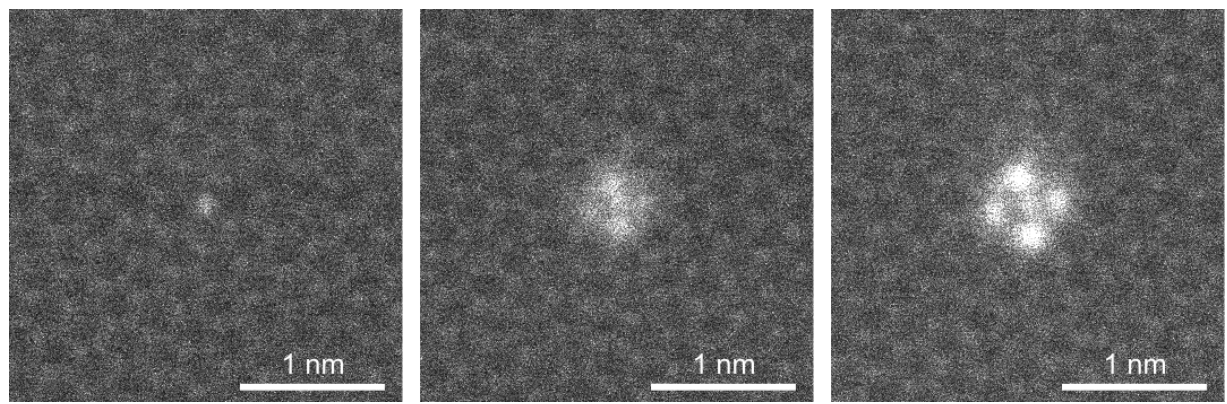

**Supplementary Figure 2.** Raw MAADF-STEM images of the structures shown in Figure 2 in the main text.

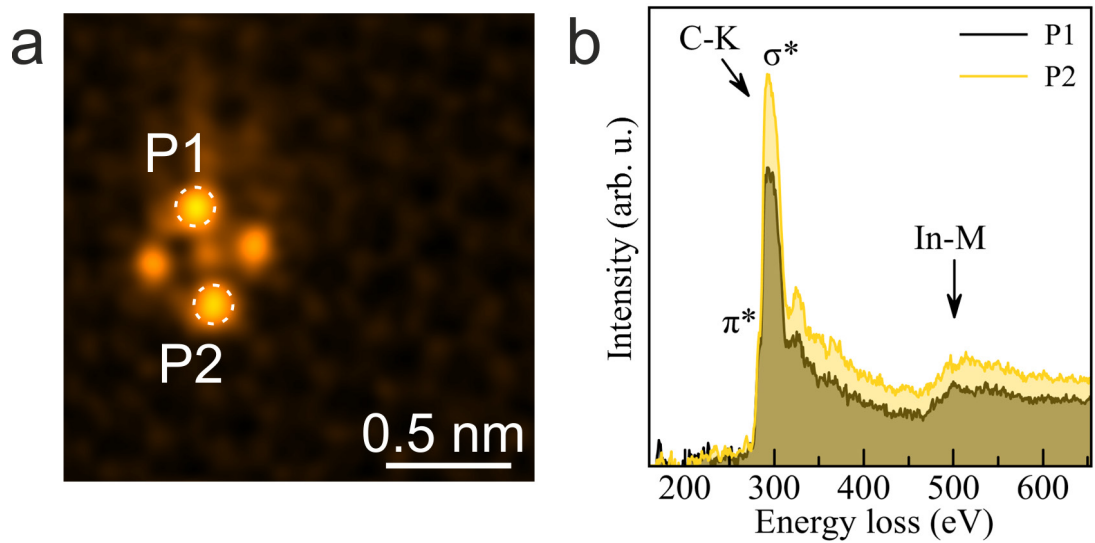

**Supplementary Figure 3.** (a) HAADF-STEM image of a 4-fold symmetric cluster (replotted from Figure 2a) and (b) EEL point spectra acquired over the brighter atoms (P1 and P2) marked by white dashed circles on panel (a). The energy dispersion is 1 eV/px and the background has been subtracted by power-law fitting.

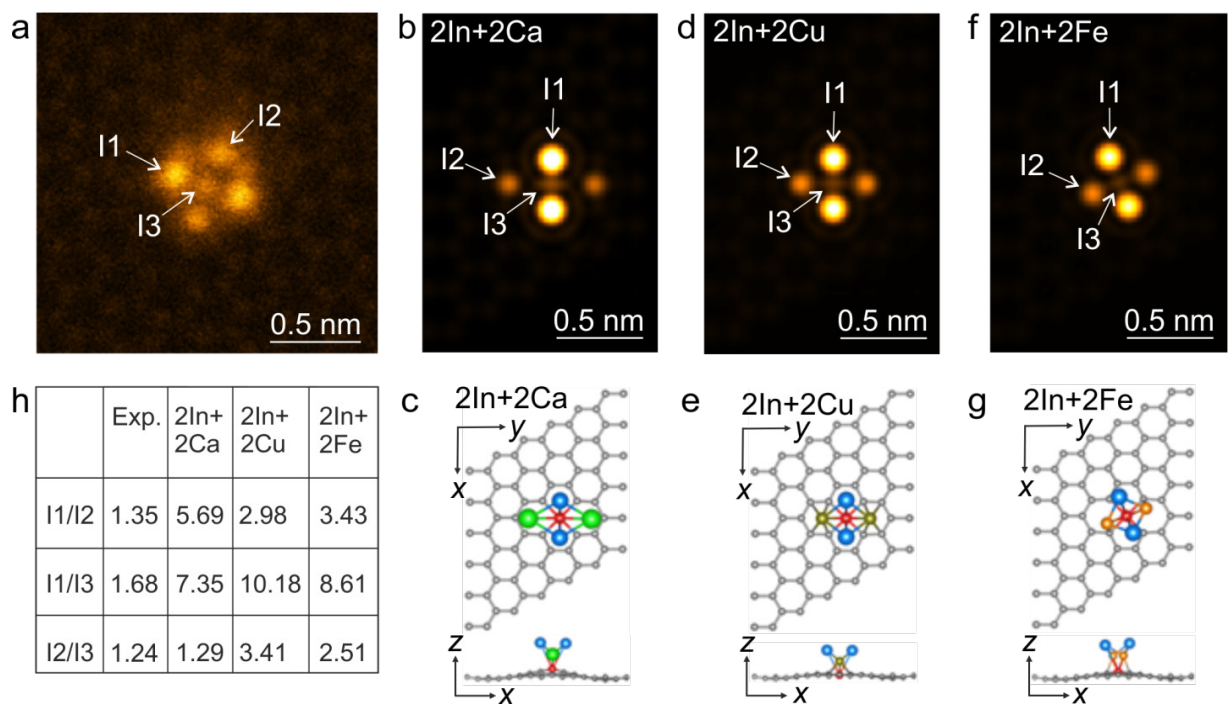

**Supplementary Figure 4.** (a) HAADF-STEM image of the hexagon-centered 4-fold symmetric In cluster shown in Figure 3a. (b) Simulated HAADF images and corresponding DFT-relaxed models for the structures consisting of (b,c) 2In+2Ca, (d.e) 2In+2Fe and (f,g) 2In+2Cu, respectively. (h) A table showing the intensity ratios measured on the atoms marked in the experimental and simulated HAADF images.

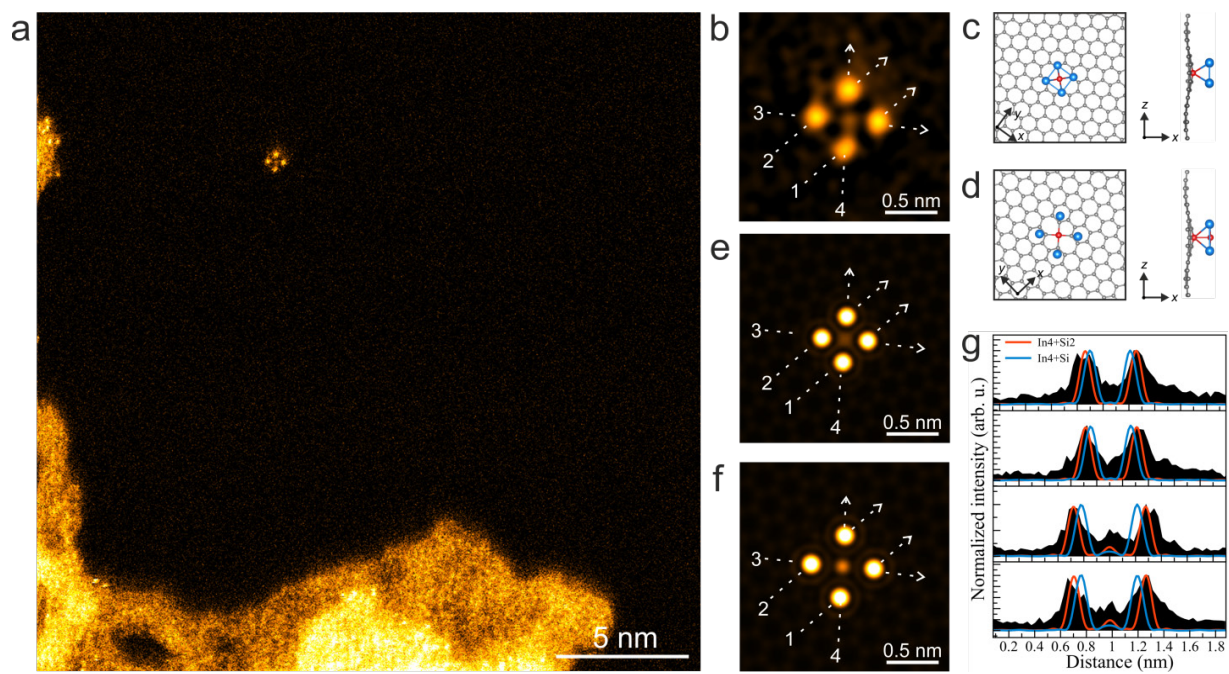

**Supplementary Figure 5**. (a) HAADF-STEM image of a 4-fold In cluster. (b) Close-up and double Gaussian filtered HAADF-STEM image of the cluster in panel a. (c) DFT-relaxed model consisting of (c) four In atoms on a 4-fold coordinated Si and (d) four In atoms and a Si atom on the 4-fold coordinated Si. (e,f) Simulated HAADF images of the models in panels c and d, respectively. (g) Intensity profiles over the white dashed lines on panels b, e and f.

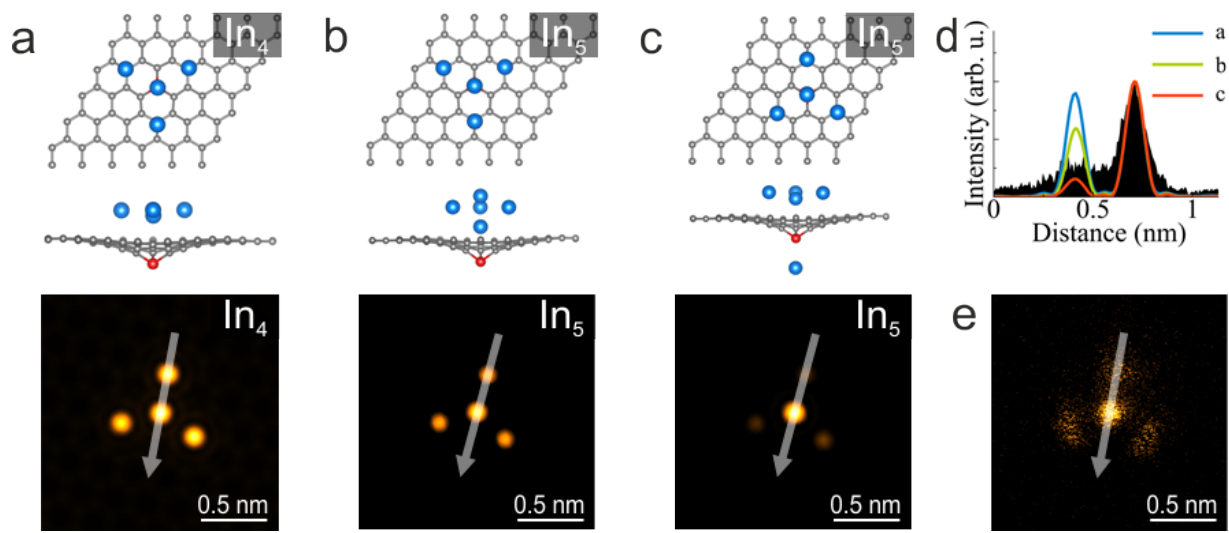

**Supplementary Figure 6.** Plan and side views of a DFT-relaxed models and corresponding simulated HAADF images showing 3-fold symmetric (a) $In_4$, (b) $In_5$ (two In atoms on top of each other at the center of cluster) and (c) $In_5$ (one In atom above and one under Si atom) clusters. (d) Intensity profiles recorded along semi-transparent white lines over simulated (a,b,c) and experimental (e) HAADF images. (e) Raw HAADF-STEM image of a 3-fold symmetric $In_5$ cluster.

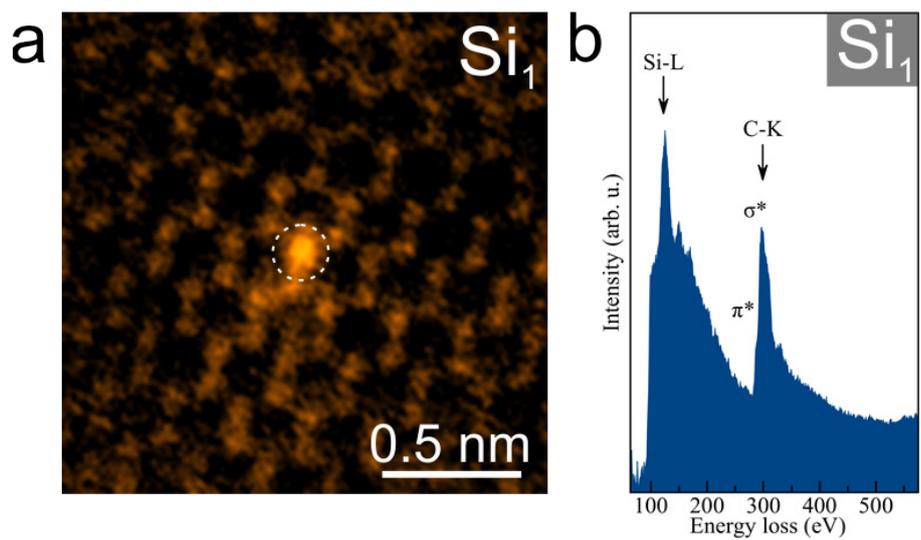

**Supplementary Figure 7.** (a) HAADF-STEM image of Si$_1$ in graphene lattice, corresponding to Figure 5j. (b) EEL spectrum acquired over the brightest atom in panel (a).

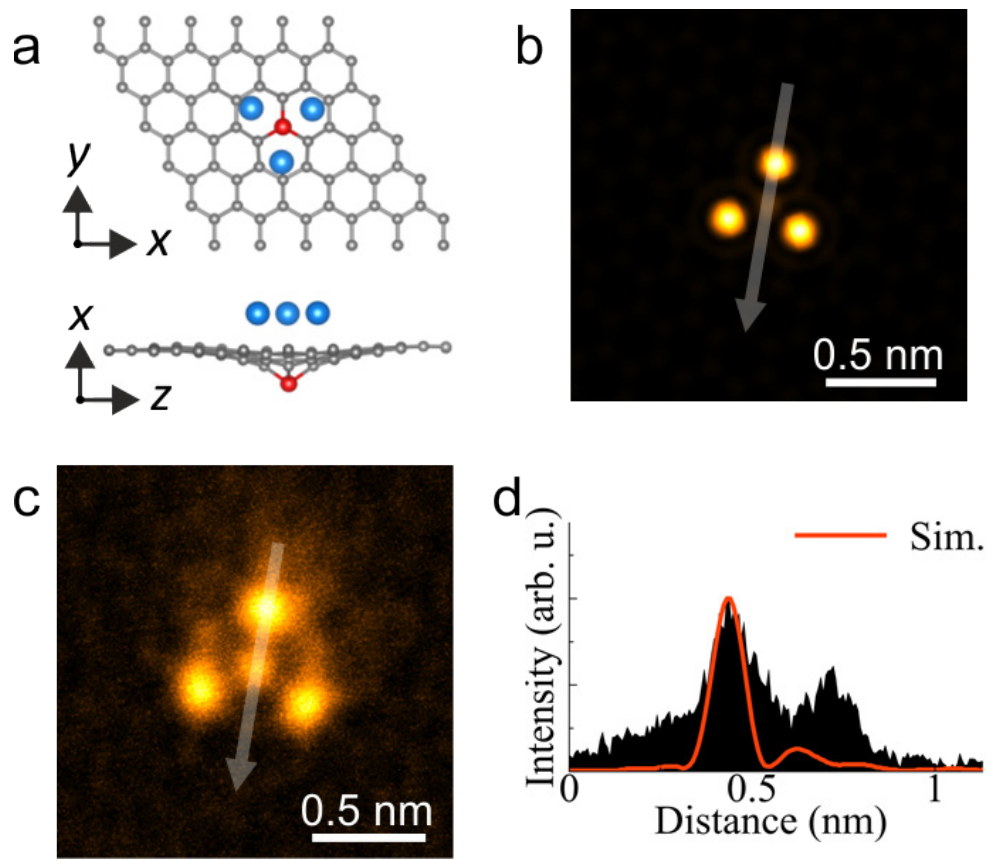

**Supplementary Figure 8.** (a) Plan and side views of a DFT-relaxed model showing 3-fold symmetric In cluster without an extra In or Si atom at the center of the cluster and (b) its corresponding simulated HAADF image. (c) HAADF-STEM image of hexagon-centered 3-fold symmetric In$_3$ cluster.

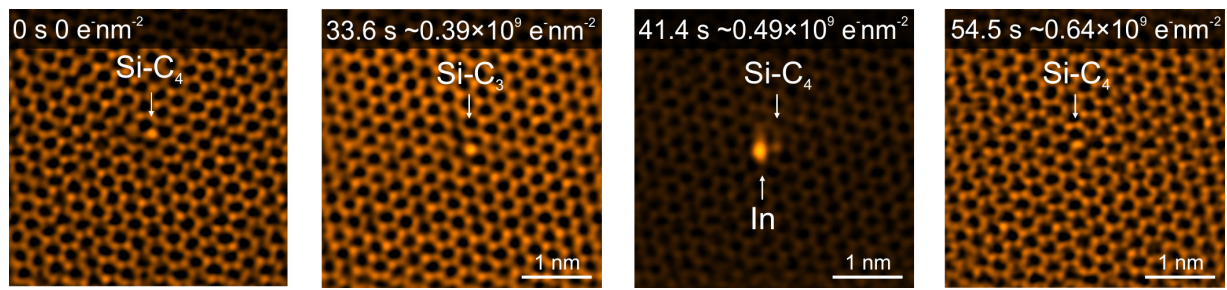

**Supplementary Figure 9.** MAADF-STEM image sequence showing a Si atom switching between 3-fold and 4-fold coordinated configurations under electron irradiation. The image acquired after 41.4 s shows an In atom briefly trapped by a 4-fold coordinated Si. Images are double Gaussian filtered after Wiener filtering.

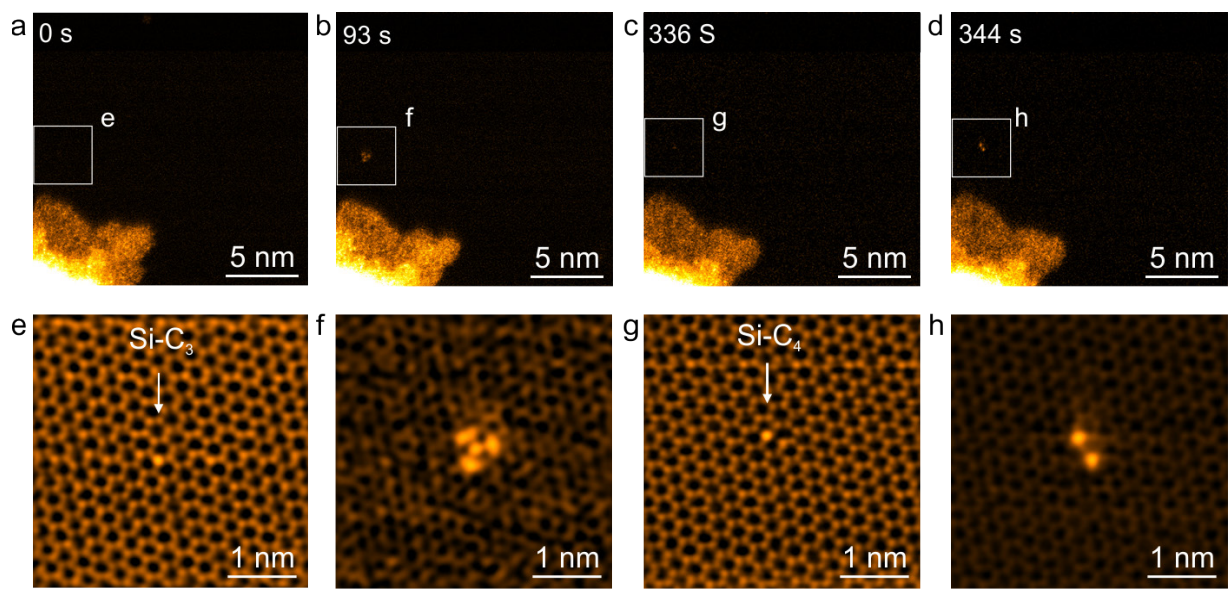

**Supplementary Figure 10.** (a-d) Large-area MAADF-STEM images. (e-f) Close up MAADF-STEM images of the areas indicated by the white frames on the images in panels a-d. The electron dose rates are $0.53\times10^6$ e$^-$nm$^{-2}$s$^{-1}$ (a-d) and $11.72\times10^6$ e$^-$nm$^{-2}$s$^{-1}$ (e-h). Images in panels a-d display raw data in false color whereas images in panels e-h are double Gaussian filtered after Wiener filtering.

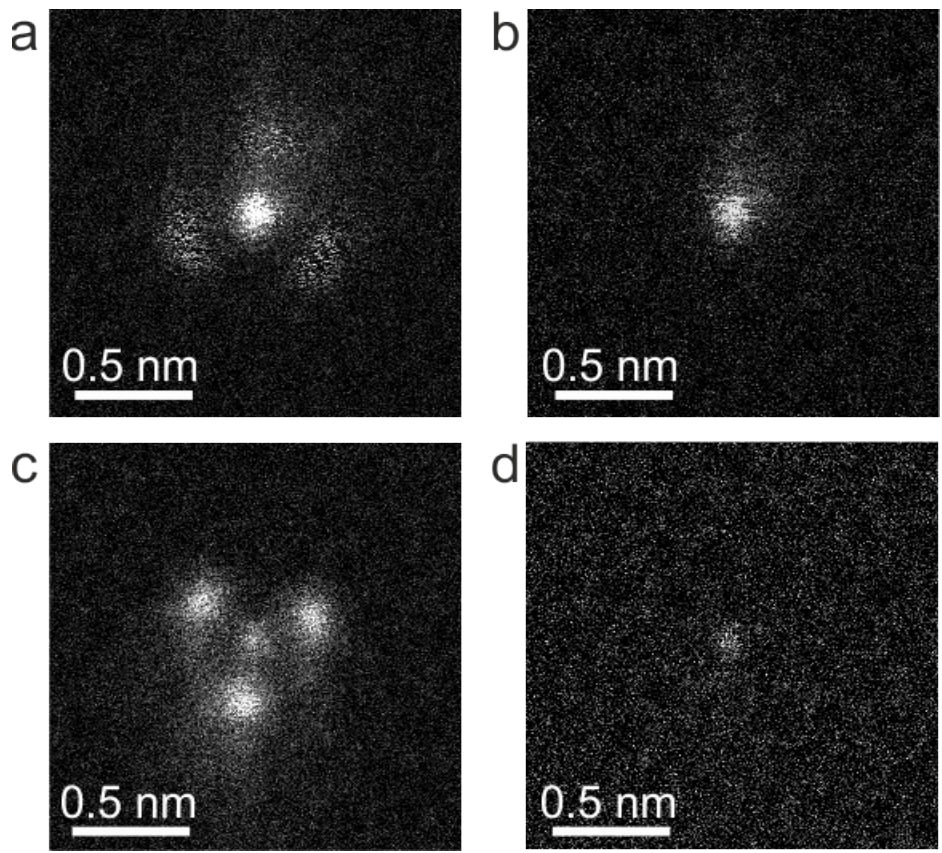

**Supplementary Figure 11.** Raw HAADF-STEM images of the structures shown in Figure 5 in the main text. (a) C-centered 3-fold symmetric $In_5$ cluster. (b) Single In atom anchored onto the Si impurity. (c) Hexagon-centered 3-fold symmetric $In_3$ cluster. (d) 3-fold coordinated Si atom in graphene.

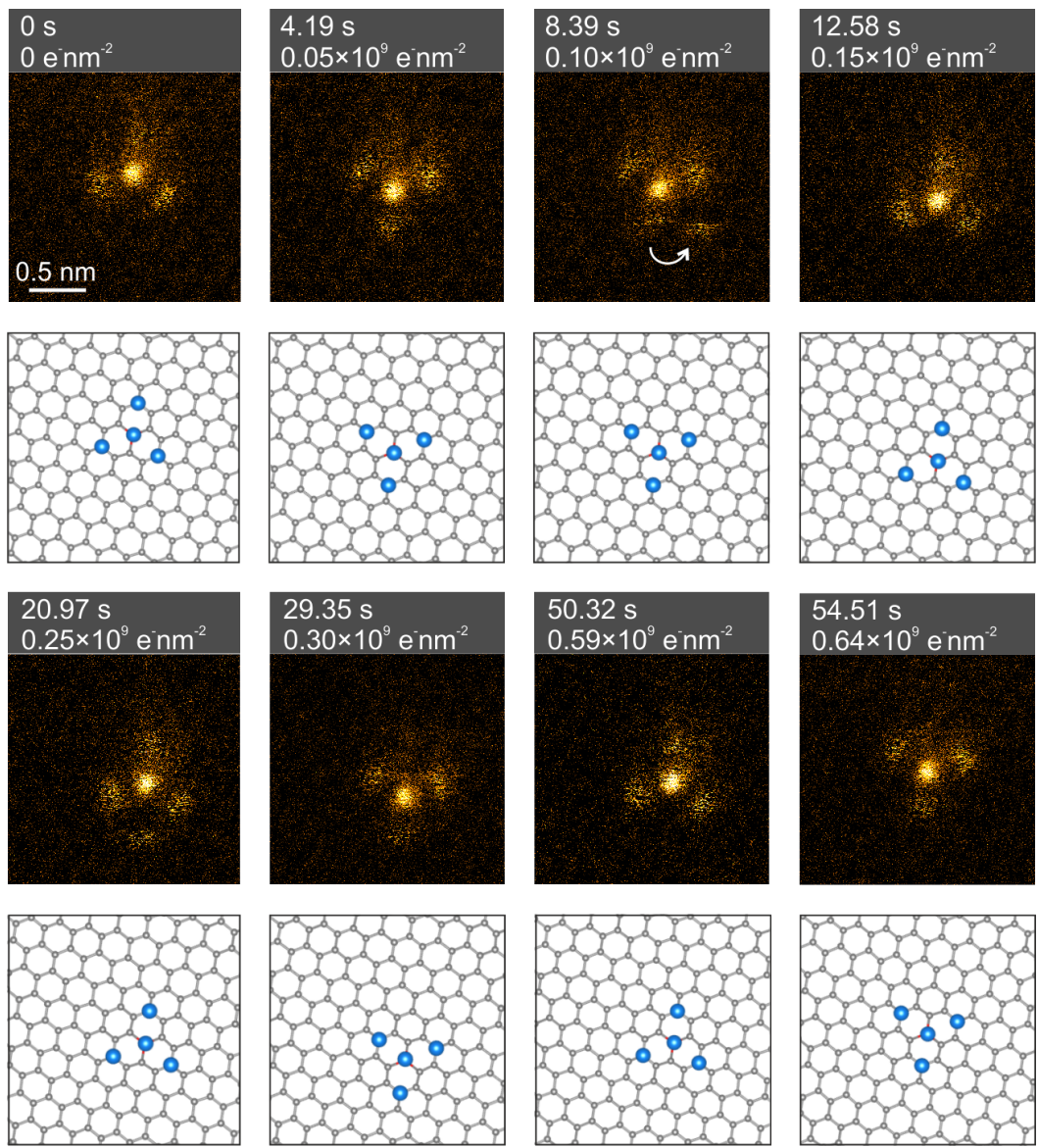

**Supplementary Figure 12.** MAADF-STEM image sequence of C-centered 3-fold symmetric In$_5$ cluster shown in Figure 6 in the main text.

**Electronic structure assessment by means of large-scale DFT**

In Supplementary Figures 13-14 we show the results of large-scale DFT calculations performed to analyse the electronic properties of the observed Si-anchored In structures on graphene in good isolation from their periodic images in the simulation. These calculations were run particularly to better understand the effect of In atom count on their spectral properties, and to estimate their relevance for applications in, e.g., single-site catalysis.

Specifically, we performed geometry relaxation and atomic population analysis using large-scale Kohn-Sham density-functional theory (DFT)[1] on 18 × 18 (~650 atom) supercells. We applied the linear-scaling DFT code ONETEP,[2–4] which uses a minimal basis of spatially-localized functions, called nonorthogonal generalized Wannier functions (NGWFs), to expand the Kohn-Sham orbitals. Accuracy equivalent to that of a plane-wave pseudopotential code was attained by variationally optimizing those Wannier functions *in situ*, separately at each geometry optimization step, in order to minimize the total energy and to refine the Hellmann-Feynman forces, including Pulay corrections,[5] used for Born-Oppenheimer geometry optimization. An out-of-plane lattice constant of 25 Å was used to separate graphene sheets from their periodic images, and an equivalent plane-wave kinetic energy cutoff of ~1020 eV was used. A common, 6.35 $a_0$ NGWF cutoff radius was applied, and no truncation of the density-matrix was applied. For each atom of a species, 4 NGWFs were allocated to C atoms, 4 to Si atoms, and 9 to In atoms, i.e. the outermost d-electrons were kept in the In valence. The PBE[6] semi-local generalized gradient exchange correlation functional was used, together with corresponding scalar-relativistic, norm-conserving pseudopotentials that we generated using the Opium code,[7] including soft non-linear core corrections. In geometry optimization, a total-energy convergence tolerance of $10^{-6}$ Ha/atom, force tolerance of $2\times10^{-3}$ Ha/Bohr, and displacement tolerance of $5\times10^{-3}$ Bohr were maintained over a convergence window of 4 geometry steps. A Gaussian smearing half-width of 0.1 eV was used in our species-decomposed Kohn-Sham density of states plots. NGWF-based Mulliken population analysis was used to partition the density, and density of states, per chemical species, and then we further divided each density of states by the number of atoms of each species to analyze how each atom contributes.

We find using Wannier-function based Mulliken atomic population analysis that for the single In atom (Supplementary Figure 13a,d), a charge of approximately 1.2 e is transferred from the graphene. This is shared between the Si atom (+0.8 e) and the In atom (+0.4 e each), with the three neighboring C atoms receiving -0.4 e and the remainder being delocalized. The optimized bond lengths in the cluster are 1.82 Å (C-Si), and 2.87 Å (Si-In). The per-atom

Kohn-Sham density of states in this cluster is dominated by near-degenerate In states distributed around the Fermi level, with further sharp In peaks spread over higher energies. This suggests that the cluster may serve as an amphoteric but more probably acceptor-like binding site, and even as a catalytic site, for small molecules.[8,9]

Referring next to the two $In_2$ dimer configurations (Supplementary Figure 13b,d for hexagon-centered and 13c,f for pentagon-centered), a larger charge of approximately 2.1 e is transferred from the graphene. This is shared between the Si atom (+0.9 e) and almost equally between the two In atoms (+0.6 e each), primarily donated by the 4 nearby C atoms (-0.4 e each). The optimized bond lengths in the hexagonal-centered cluster are, on average 1.89 Å (C-Si), 3.04 Å (Si-In), 4.01 Å (In-In). The optimized bond lengths in the pentagon-centered cluster are, on average 1.90 Å (C-Si), 2.94 Å (Si-In), 3.65 Å (In-In). The per-atom Kohn-Sham density of states for both of these clusters is characterized by a half-filled four-fold-degenerate peak at the Fermi level, of predominantly In character. A dense group of In levels is spread over higher energies.

In the $In_4$ cluster (Supplementary Figure 14a-c), a charge of approximately 1.9 e is transferred from the graphene. This is shared between the Si atom (+0.8 e) and almost equally between the four In atoms (+0.3 e each), and is primarily donated by 4 nearby C atoms (-0.4 e each). The optimized bond lengths in the cluster are, on average 1.95 Å (C-Si), 3.30 Å (Si-In), 2.75 Å (In-C), with the In-In bond lengths coming in pairs of length 3.07 Å and 3.14 Å. Amongst those studied, this cluster exhibits the most prominent partially-filled In peak at the Fermi level.

Finally, in the $In_6$ cluster (Supplementary Figure 14d-f), approximately 1.7 e is transferred, but now the only significantly charged atoms are the Si (+0.8 e), the 2 lowest-lying In atoms adjacent to Si (+0.3 e), and the 4 C atoms bound to Si (-0.3 e). The cluster structure is more complex, comprising pairs of In atoms at three different altitudes with respect to the graphene surface, with the atoms in the first and third rows being aligned close to vertically with respect to each other, above the center of the 5-member C-Si ring. The second row sits above the center of the 6-member C-Si ring. The intra-row In-In bond length varies, from first to third row, as 3.24 Å, 4.33 Å, 4.14 Å. The other salient bond lengths are 1.92 Å (C-Si), 3.30 Å (Si-In), 2.91 Å (Si-In first row), 2.93 Å (In-In first row to third row). The first-row In atoms are sometimes closer to the C atoms in the 5-member C-Si ring than to the Si atom, at distances of approximately 2.8-2.9 Å. The per-atom Kohn-Sham density of states in this cluster again reveals a Fermi density dominated by In states, albeit not as sharply so as in the 4-atom case and, for a given energy, not equally among the In rows. Again, the charge-transfer to the cluster from the graphene serves to pin the bottom of the In 5p-like density of states to the Fermi level, potentially providing overall a relatively spatially-localised but prominent conduit for further charge transfer and possible associated catalytic activity.[10–12]

On the basis of these calculations alone it is of course impossible to judge which of these clusters may provide the best activity and selectivity for a given chemical reaction, if any are indeed suitable for single-site catalysis. However, we can say that all of them should provide a strongly polarizing localized environment, and most likely act as electron acceptors, possibly amphoterically. A range of In-In and In-Si bond lengths are available across the clusters studied, and shorter bond lengths may suggest greater resistance to catalyst erosion and consumption. Overall, these clusters present an interesting avenue to extent the concept of the single-atom catalyst to more complex cluster geometries.

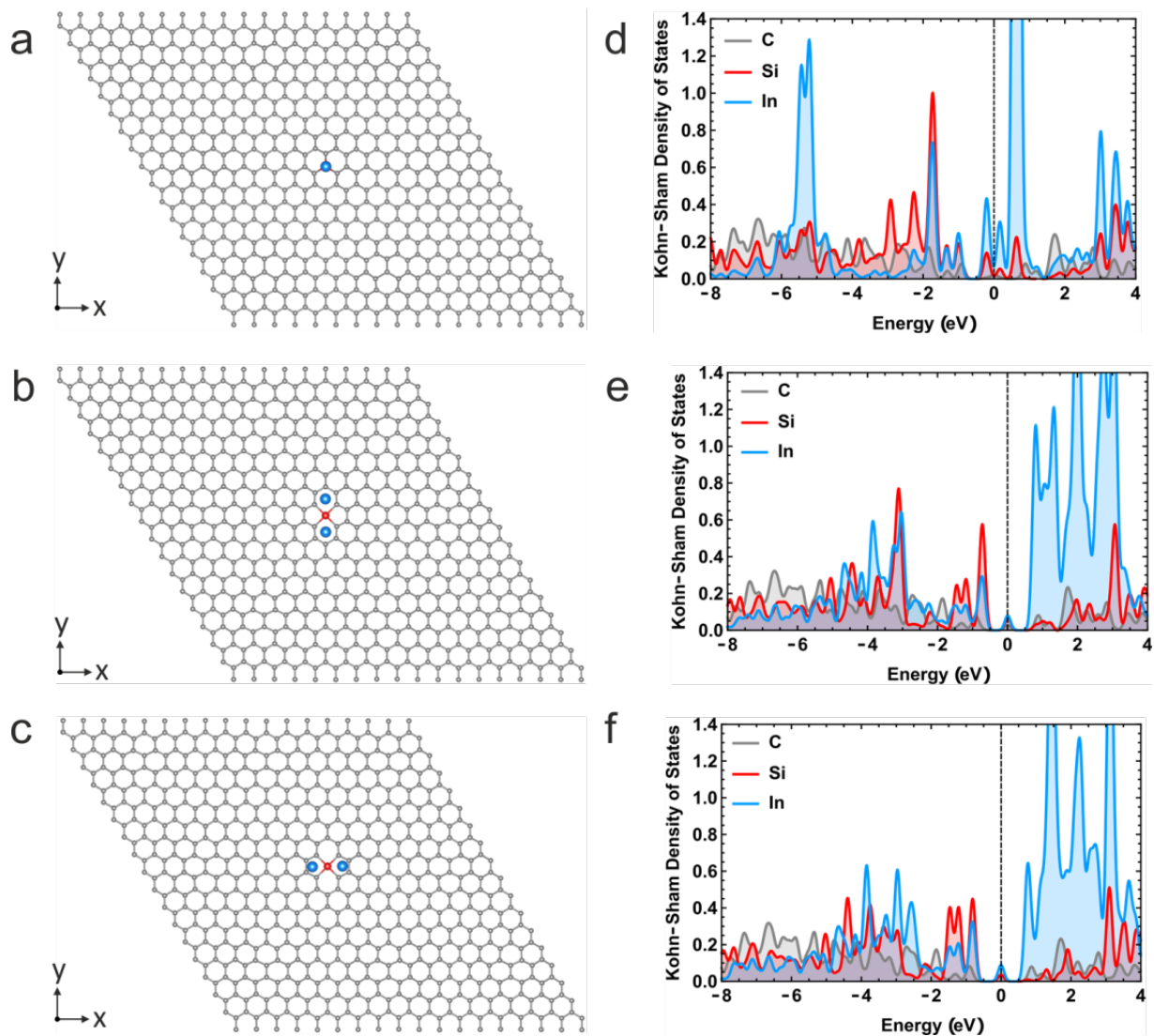

**Supplementary Figure 13.** DFT-relaxed models showing 18 × 18 graphene cells containing (a) single In, (b) In$_2$ (hexagon-centered) and (c) In$_2$ (pentagon-centered) clusters. (d-f) Wannier-function based local density of Kohn-Sham states for single In atom, hexagon-centered In$_2$ dimer and pentagon-centered In$_2$ dimer, respectively, renormalized per atom.

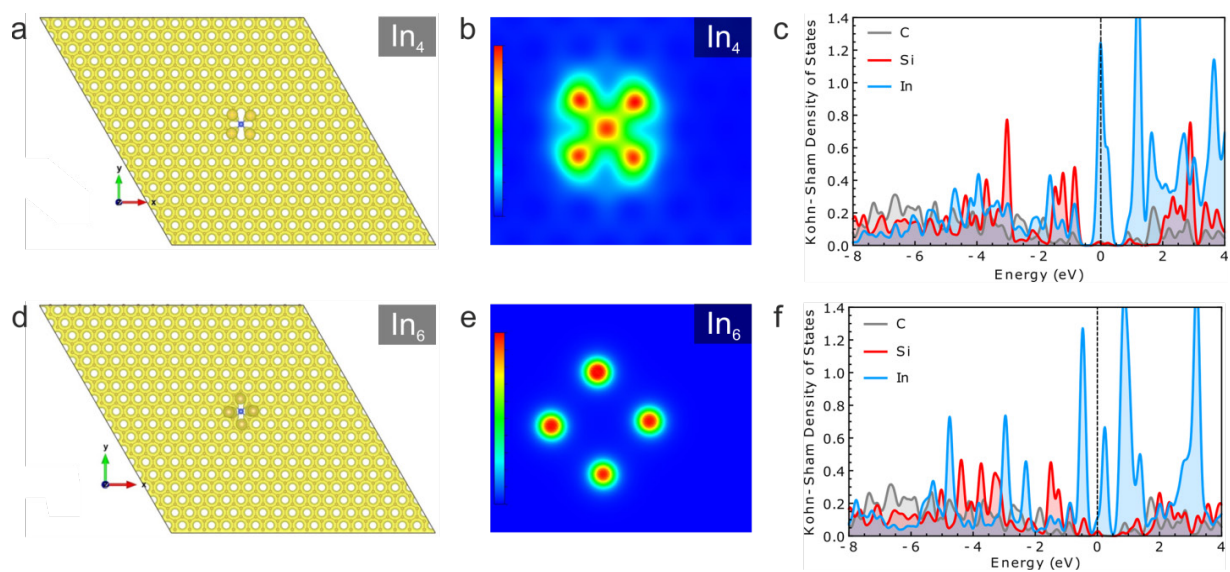

**Supplementary Figure 14.** The electron density and Kohn−Sham density of states. (a,d) Isosurfaces of the calculated charge densities for 18 × 18 graphene cells containing 4-fold symmetric $In_4$ and $In_6$ clusters. (b,e) Contour plots showing zoomed-in planar cross-sections of the charge densities of 4-atom and 6-atom clusters, where the colour scale is set close to the minimum and maximum values on each plane. In the 4-atom case in (b), this cross-section is taken on a plane parallel to the graphene, equidistant between the Si atom and the 4 In atoms. In the 6-atom case in (e), the plane is instead set at an altitude above graphene that is equidistant between the second and third In rows. (c,f) Wannier-function based local density of Kohn-Sham states for 4-atom and 6-atom 4-fold symmetric clusters, respectively, renormalized per atom.

**Supplementary Videos**

Supplementary Videos are available from the authors upon request.

**Supplementary References**